\documentstyle[11pt,epsfig]{article}
\def\be{\begin{equation}}
\def\ee{\end{equation}}
\def\bea{\begin{eqnarray}}
\def\eea{\end{eqnarray}}
\def\ben{\begin{enumerate}}
\def\een{\end{enumerate}}
\def\etal{{\it et. al.}}
\def\del{\partial}
\def\np {Nucl. Phys.}
\def\prl{Phys. Rev. Lett.}
\def\text#1{{\rm #1}}

\def\Lag{{\cal L}}

\def\e{\mbox{e}}

\def\hatc{{\hat c}}

\def\vecq{{\vec q}}
\def\vecp{{\vec p}}
\def\veck{{\vec k}}
\def\vecs{{\vec \sigma}}

\begin{document}
\begin{center}
\begin{flushright}
SNUTP 97-047
\end{flushright}
\vskip 1.0cm
{\Large\bf In-Medium Effective Axial-Vector Coupling Constant}
\vskip 1cm
{\large Tae-Sun Park\\
Department of Physics, Hanyang University, Seoul 133-791, Korea\\
 Hong Jung\\
Department of Physics, Sookmyung Women's University, Seoul 140-742, Korea\\
 Dong-Pil Min\\
Department of Physics and Center for Theoretical Physics,\\ 
Seoul National University, Seoul 151-742, Korea}
\end{center}
\vskip 2cm

The axial-vector coupling constant
in nuclear medium is systematically studied using chiral
perturbation theory. For normal nuclear matter,
the first non-trivial corrections are estimated based on
the independent particle approximation taking into account
nuclear correlations in nuclear medium.
The one-pion-exchange (OPE) two-body interactions
turn out to
enhance the coupling constant by $\delta g_A/g_A\simeq (0.09 \sim 0.15)$.
The role of the four-nucleon contact interaction, 
whose low-energy constants are yet to be determined, 
is analyzed in connection with the empirical quenching factor of
$\delta g_A/g_A\simeq -0.2$ in the presence of the 
nuclear short-range correlation. 
Conventional approaches are discussed in the light of our result.

\newpage
The nucleon's axial-vector coupling constant in nuclear medium $g_A^*$
is empirically known to be quenched by the amount of
$\frac{\delta g_A}{g_A}\equiv \frac{g_A^*-g_A}{g_A}\simeq -0.2$\cite{exp}.
Since the in-medium quenching is very important for various places
in nuclear physics,
there have been many efforts for its explanation\cite{theory}.
One well-known example is the question whether or not 
the pion condensation is practically possible 
at a low nuclear density\cite{pioncond}.
Indeed, wide discussions on the change of matter properties 
at the extreme condition take place in conjunction with 
CERN-CERES, RHIC and CEBAF experiments.

In most conventional approaches, a major contribution to the quenching of
the axial-vector coupling in nuclear medium is from the renormalization 
due to the $\Delta$-hole screening induced by the short-range interaction 
in $NN\leftrightarrow N\Delta$ (Hartree) channel with the net contribution 
from other mechanisms presumed to be negligible due to the mutual 
cancellation and/or due to the famous
nuclear correlation\cite{rho1}. Also, some fascinating efforts
have been given to figure out the medium effects on physical quantities
by taking the tree order only with some scaled parameters of 
chiral Lagrangian, which is known as
``Brown-Rho scaling"\cite{BR}. 
We wish to understand the conventional approaches
based on a well-defined systematic method.

In this letter, we aim to study the nucleon's axial-vector coupling 
in nuclear medium in a systematic way using chiral perturbation theory 
generalized to many nucleon systems\cite{Weinberg, pmr1}. 
Chiral perturbation theory has been successful in explaining the nuclear
force\cite{Kolck1} 
and the so-called chiral filter 
phenomena\cite{pmr1, KDR, TSP}. 
In order to extract the axial-vector coupling constant, 
the relevant quantity is the transition amplitude of the 
Gamow-Teller operator, that is, the space part of the
axial-vector current in the limit $k^\mu\rightarrow 0$
where $k^\mu$ is the momentum carried by the external field.
For a single nucleon in free space, the Gamow-Teller operator is given as
\be
g_A\, \tau^{\pm} {\vec \sigma}.
\ee
In nuclear medium, the axial-vector current is modified by the presence of
other nucleons, and how the transition amplitude of the Gamow-Teller operator
changes is what we are interested in.

The chiral power counting rule for $A$-nucleon processes\cite{Weinberg} is 
that for a Feynman graph with $V_i$ vertices of type $i$, $L$ loops, 
and $C$ separately connected pieces,  
the number of powers of $Q$ which is the typical low momentum scale
is given as                             
\begin{equation}
\nu = 4-A-2C+2L +\sum_{i}V_i\Delta_i 
\;\;{\rm with}\; \Delta_i=d_i+{n_i\over 2}-2,
\end{equation}
where $n_i$ is  the number of nucleon lines and 
$d_i$ is the number of derivatives or powers of $m_{\pi}$ 
at the $i$-type vertex.
Chiral symmetry guarantees $\Delta_i\ge 0$\cite{Weinberg}, 
so the effective Lagrangian can be ordered according to $\Delta_i$: 
\begin{equation}
{\cal L}_{eff} = {\cal L}_0 + {\cal L}_1 +{\cal L}_2 +\cdots,
\end{equation}
where ${\cal L}_n$ has vertices with $\Delta_i=n$.
However, in the presence of an external field, 
the lower bound on $\Delta_i$ is modified to $\Delta_i\ge -1$ 
because of the gauge invariant derivative coupling\cite{rho2}.

In this work, we take the nucleon and the pion  as pertinent degrees of 
freedom. 
The contribution of $\Delta$ will be properly encoded
in some of the low-energy constants.
Then, the leading order Lagrangian is
\bea
\Lag_0 &=& {\bar B}\left[ i v\cdot D
+ 2 i g_A S\cdot \Delta \right] B
- \frac{1}{2} \sum_A C_A \left({\bar B} \Gamma_A B\right)^2
\nonumber \\
&&+\, f_\pi^2 {\rm Tr}\left(i \Delta^\mu i \Delta_\mu\right)
+ \frac{f_\pi^2}{4} {\rm Tr}(\chi_+)
\label{chiralag2} \eea
with                                                                               
\bea
D_\mu N &=& (\del_\mu + \Gamma_\mu) N ,
\nonumber \\
\Gamma_\mu &=& \frac{1}{2} \left[\xi^\dagger,\, \del_\mu \xi\right]
-\frac{i}{2}\xi^\dagger {\cal R}_\mu \xi - \frac{i}{2}
\xi {\cal L}_\mu\xi^\dagger,
\nonumber \\
\Delta_\mu &=& \frac{1}{2} \left[\xi^\dagger,\, \del_\mu \xi\right]
+\frac{i}{2}\xi^\dagger {\cal R}_\mu \xi - \frac{i}{2}
\xi {\cal L}_\mu\xi^\dagger,
\nonumber \\
\chi_+ &=& \xi^\dagger \chi \xi^\dagger + \xi \chi^\dagger \xi
\label{deltamu}\eea
where 
${\cal R}_\mu = \frac{\tau^a}{2} \left(
  {\cal V}^a_\mu + {\cal A}^a_\nu\right)$
and
${\cal L}_\mu = \frac{\tau^a}{2} \left(
  {\cal V}^a_\mu - {\cal A}^a_\nu\right)$
denote the external gauge fields, 
$\chi$ is proportional to the quark mass matrix
and it becomes, if we neglect the small isospin-symmetry breaking,
$\chi=m_\pi^2$ in the absence of the external scalar 
and pseudo-scalar gauge field, and 
\be          
\xi = \sqrt{\Sigma} = 
  {\rm exp}\left(i\frac{{\vec \tau}\cdot {\vec \pi}}{2 f_\pi}\right).
\ee
For convenience, we will work in a reference frame in which 
the four velocity $v^{\mu}$ and the spin operator $S^{\mu}$ are 
\be
v^\mu = (1,\, {\vec 0})
\ \ \ 
\mbox{and} \ \ 
S^\mu = \left(0,\, \frac{{\vec \sigma}}{2}\right).
\ee 
{}From \cite{meissner} and \cite{Kolck3}\footnote{
Our definition of the pion field is different from that used in Ref. 
\cite{Kolck3}: We should attach a minus sign for the pion field of 
Ref.\cite{Kolck3} and then make the interaction form of Ref.\cite{Kolck3} 
manifestly Lorentz-invariant and chiral-invariant.},  
the next-to-leading order Lagrangian 
including the four-fermion contact terms
can be written as
\bea
\Lag_1 &=& {\bar B} \left(
  \frac{v^\mu v^\nu - g^{\mu\nu}}{2 m_N} D_\mu D_\nu
   + 4 c_3 i\Delta\cdot i\Delta
   + \left(2 c_4 + \frac{1}{2m_N}\right)
  \left[S^\mu, \,S^\nu\right] \left[ i \Delta_\nu,\, i\Delta_\nu\right]
\right)B
\nonumber \\
 && -\ 4 i d_1 \,
 {\bar B} S\cdot \Delta B\, {\bar B} B
 + 2 i d_2 \,
  \epsilon^{abc}\,\epsilon_{\mu\nu\lambda\delta} v^\mu \Delta^{\nu,a}
 {\bar B} S^\lambda \tau^b B\, {\bar B} S^\delta \tau^c B
+ \cdots,
\label{Lag1}\eea
where $\epsilon_{0123}=1$, $\Delta_\mu = \frac{\tau^a}{2} \Delta^a_\mu$, 
and only those terms relevant for our study are  shown explicitly.

The leading contribution to the transition amplitude of the Gamow-Teller
operator is given by the one-body currents ($L=0$, $C=A$) 
with one $\Delta_i=-1$ interaction ($\nu=3-3A$).
The one-body currents, including corrections to all order, 
are entirely determined by the experiment,
$g_A\simeq 1.26$. 

The first corrections due to the nuclear medium are given by tree graphs of
two-body currents ($L=0$, $C=A-1$) with one $\Delta_i=-1$ interaction and 
interactions with $\Delta_i=0$ ($\nu=5-3A$).
Since there are no $\Delta_i=-1$ four-nucleon contact interactions,
at order $\nu=5-3A$ we have only one-pion-exchange
graphs, but they do not contribute to
the Gamow-Teller operator because
the currents from one-pion-exchange graphs 
are proportional to $v^\mu=(1, \, {\vec 0})$.
Thus, the leading order medium corrections to the
space part of the axial-vector currents are kinematically
suppressed. 
Therefore the higher order contributions
including the vertex corrections as well as
the short-range contributions
become important\cite{KDR}.
This situation is contrary to the case of the axial-charge
in which  the one-body currents are suppressed
while the leading two-body currents are not\cite{pmr1,TSP}.

\begin{figure}
\centerline{\epsfig{file=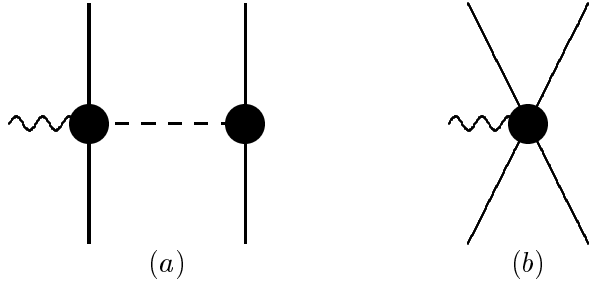}}
\caption{Generic graphs contributing to the exchange current:
the graph $(a)$ is the OPE graph
and the graph $(b)$ is the four-fermion contact graph.
The large filled circles
denote the one-nucleon and one-pion irreducible graphs
and the solid (dashed) lines
the renormalized nucleon (pion) propagators.
The renormalization has been taken up to
$1/m_N$ correction.
}
\label{GeneV}
\end{figure}
The first non-trivial medium corrections come from tree-graphs of 
two-body currents with $\Delta_i=0$ interactions ($\nu=6-3A$).
More specifically, we can have one-pion-exchange (OPE) graphs (Fig.~1$(a)$) 
and four-nucleon contact graphs (Fig.~1$(b)$).
In the limit when momentum carried by external field goes to zero,
the space part of the axial-vector currents reads
\bea
{\vec J}_5^\pm(\vec q) &=&
- \frac{g_A}{m_N f_\pi^2}\, \frac{1}{m_\pi^2 + \vecq^2}
 \left\{
 \frac{i}{2} (\tau_1\times\tau_2)^\pm
    ({\vec {\bar p}}_1\,\vecs_2 \cdot \vecq 
       + {\vec {\bar p}}_2\,\vecs_1 \cdot \vecq)
 \right.
 \nonumber \\
 &&\ \ \ +\ 
  2 \hatc_3 \vecq 
  (\tau_1^\pm \vecs_1\cdot \vec q + \tau_2^\pm \vecs_2\cdot \vec q)
 \nonumber \\
 &&\left.\ \ \ +\ 
 \left(\hatc_4 + \frac14\right) (\tau_1\times\tau_2)^\pm
 \left(\vecs_1\times \vecq \, \vecs_2\cdot \vecq 
 -\vecs_2\times \vecq \, \vecs_1\cdot \vecq \right)
 \frac{}{}\right\} 
\nonumber \\
 &-& \frac{2 g_A}{m_N f_\pi^2} \left[
  \hat d_1 (\tau_1^\pm \vecs_1 + \tau_2^\pm \vecs_2)
 - \hat d_2 (\tau_1\times\tau_2)^\pm (\vecs_1\times \vecs_2)
 \right]
\eea
where 
${\vec q} \equiv {{\vec p}_2}' - {\vec p}_2$,
the subscript $l$ of $\tau_l^\pm=\frac12 (\tau_l^1 \pm i \tau_l^2)$, 
${\vec \sigma}_l$, 
and ${\vec {\bar p_l}} \equiv \frac12 ({\vec p}_l + {{\vec p}_l}')$
is the particle index,
$l=1$ for the first nucleon and $l=2$ for the second nucleon,
and we have defined dimensionless parameters 
$\hatc_i$'s and $\hat d_i$'s by
\be
c_i \equiv \frac{1}{m_N} \hatc_i \ \ \ 
\mbox{and}\ \ \
d_i \equiv \frac{g_A}{m_N f_\pi^2} \hat d_i\,.
\label{c-normal}\ee
The first term in the curly bracket comes from
$\frac{1}{2 m_N} {\bar B} {\vec D}^2 B$ of eq.(\ref{Lag1})
and will be referred to as the kinetic term.
The low-energy constants $c_3$ and $c_4$ have been 
determined from the experiment 
by Bernard \etal\cite{meissner}:
\bea
c_3 &=& -5.29 \pm 0.25\ \mbox{GeV}^{-1},
\nonumber \\
c_4 &=&  3.63 \pm 0.10\ \mbox{GeV}^{-1}.
\label{c3c4}\eea
These values of $c_3$ and $c_4$ are almost saturated by the 
resonance-exchange contributions:
\bea
c_3^{\rm Res} &=& c_3^\Delta + c_3^S + c_3^R
  = \left(-3.83 -1.40 - 0.06\right) \ \mbox{GeV}^{-1}  
  = -5.29 \ \mbox{GeV}^{-1} , \nonumber \\
c_4^{\rm Res} &=& c_4^\Delta + c_4^\rho + c_4^R
  = \left(1.92 + 1.63 + 0.12\right) \ \mbox{GeV}^{-1} 
  = 3.67 \ \mbox{GeV}^{-1}
\eea
where the superscripts $\Delta$, $\rho$, $S$  and  $R$
denote the contributions from the exchange of
$\Delta(1232)$, $\rho$-mesons, scalar-mesons
and the Roper resonance, respectively.
In terms of the dimensionless constants $\hat c_3$ and $\hat c_4$,
the above eq.(\ref{c3c4}) gives us
\bea
\hatc_3 &\simeq& - 4.97 \pm 0.23,
\nonumber \\
\hatc_4 &\simeq& 3.41 \pm 0.09 .
\label{hat_c}\eea
For the low-energy constants $\hat d_1$ and $\hat d_2$, 
however, we do not have any empirical information yet. 
So we simply take these constants as free parameters in this work.

Now we are in the position to calculate the matrix elements of the
space part of the two-body axial-vector currents. First we estimate the matrix
elements using the independent particle approximation,
{\it i.e.}, we  take the initial (final) wavefunction as the simple
product of one-nucleon state $\alpha$ ($\beta$) and the Fock
state $|F\rangle$ which describes the Fermi sea,
$$ | F\rangle = \prod_{h \in F} a^\dagger_h |vac\rangle,$$
where $h$ denotes a state in the Fermi sea.
Since the momentum carried by external field is zero,
only those states on the Fermi surface can contribute to the matrix elements
and we have
\be
\vecp_\beta = \vecp_\alpha \equiv \vecp \;\; {\rm with}\;\;
\ \ \ |\vecp|=k_F .
\ee
Then the two-body matrix elements relevant for the 
axial-vector coupling constant 
are
\be
\langle \beta; F | J^{i,\pm}_{5}(\vecq) | \alpha; F\rangle
= \sum_{h \in F} \left[
\langle \beta, h | J^{i,\pm}_5(\vecq) | \alpha, h \rangle
- \langle \beta, h | J^{i,\pm}_5(\vecq) | h, \alpha \rangle
\right],
\ee
where the first term is the direct (or Hartree) term and the second
is the exchange (or Fock) term.
We average over the the direction of the probed nucleons
in order to obtain the in-medium axial-vector coupling constant, 
\be
\int\frac{d\Omega_{\hat p}}{4\pi} 
\langle \beta; F | J^{i,\pm}_{5}(\vecq) | \alpha; F\rangle
\equiv \delta g_A\, \tau_{\beta\alpha}^\pm \sigma_{\beta\alpha}^i,
\ \ \
\delta g_A = \delta^H g_A + \delta^F g_A
\ee
where
$\tau^\pm_{\beta\alpha}\equiv \langle t_\beta | \tau^\pm | t_\alpha\rangle$,
$\sigma^j_{\beta\alpha}\equiv 
\langle s_\beta | \sigma^j | s_\alpha\rangle$ with
$t_{\alpha}$($t_{\beta}$) and $s_{\alpha}$($s_{\beta}$) being the 
quantum numbers for the third components of the isospin and the 
spin of a state $\alpha$($\beta$) respectively, and
we have separated the Hartree contribution ($\delta^H g_A$)
from the Fock contribution ($\delta^F g_A$).
Note that the Hartree term for the OPE contribution is 
identically zero because it involves only $\vecq= {\vec 0}$ 
and the corresponding axial-vector current vanishes, thus we have
\be
\frac{\delta^H g_A}{g_A} =
   - 2 \hat d_1 \frac{\rho}{m_N f_{\pi}^2} 
\label{gAHartree}
\ee
with $\rho=\frac{2k_F^3}{3\pi^2}$ being the density of 
symmetric nuclear matter.
It is also quite straightforward to evaluate the Fock contribution,
\be
\frac{\delta^F g_A}{g_A} =
\frac{4}{m_N f_{\pi}^2} 
\int^{k_F} \frac{d^3 \veck}{(2\pi)^3}\,
\left[\frac{{1\over 24}(\vec k^2 - \vec p^2) 
    +{1\over 3}(\hat c_3- 2 \hat c_4-\frac12)(\vec k - \vec p)^2}
             {m_\pi^2 + (\vec k - \vec p)^2} + \hat d_1 + 2 \hat d_2
  \right] .
\label{M1}
\ee
Performing the momentum integral over the Fermi sea, we have
\be
\frac{\delta^F g_A}{g_A} =
        \frac{\rho}{m_N f_{\pi}^2} \left\{
\frac{k_F^2}{6 m_\pi^2} H_1 \left(\frac{k_F}{m_\pi}\right)
 + \frac{1}{3} \left(\hat c_3- 2\hat c_4 +\frac14 \right) 
 \left[ 1 - H_0 \left(\frac{k_F}{m_\pi}\right)\right]
 + \hat d_1+ 2 \hat d_2\right\}
\label{M2}
\ee
with
\begin{eqnarray}
H_0(x) &=&  \frac{3}{2 x^2} \left[ 1 - \frac{1}{x} \tan^{-1}(2x)
 +\frac{1}{4 x^2}\log(1+4 x^2)\right],\\
H_1(x) &=& \frac{3}{8 x^4} \left[ 1 + 2 x^2
 +\left(1 + \frac{1}{4 x^2}\right)\log(1+4 x^2)\right].
\end{eqnarray}
For normal nuclear matter, $k_F\simeq 1.36\;{\rm fm}^{-1}$, 
$\rho\simeq 0.17\;{\rm fm}^{-3}$, thus leading to
$H_0(\frac{k_F}{m_\pi}) \simeq 0.20$ and
$\frac{k_F^2}{m_\pi^2} H_1(\frac{k_F}{m_\pi}) \simeq 0.55$.

\begin{table}
\begin{center}
\begin{tabular}{|c||rrrrr|r|} \hline
 & $\Delta$ &  $\rho$ & $S$ & $N^*$ & kin. & Experiment \\ \hline
Zero-range &  $-0.386$ & $-0.164$ & $-0.070$ & $-0.015$ & $-0.013$ &
 $-0.672$ \\
Finite-range &  0.078 & 0.033 & 0.014 & 0.003 & 0.018 & 
 $0.151$ \\ \hline
Sum &
 $-0.308$ & $-0.131$ & $-0.056$ & $-0.012$ & $0.004$ & 
 $-0.521$ \\ \hline
\end{tabular}
\end{center}
\caption\protect{One-pion-exchange (Fig.~1$(a)$) contributions to the
$\delta g_A / g_A$.
The contributions from the resonance-exchanges through the $\pi{\cal A} NN$
vertex has been listed in column $2 \sim 5$. The last column represents
all the contributions with $\hatc$'s 
fixed from the
$\pi N$ scattering data\cite{meissner}
and the kinetic term.
The second row shows zero-ranged (in position space) contributions,
that is, those which are constant in momentum space,
the third row does finite-ranged contributions and the last row
does the sum of the two.}
\end{table}

We now discuss the numerical outcome on the quenching factor $\delta g_A/g_A$, 
though {\it naive}, firstly without taking the correlation 
into account.
In Table~1, we have listed the contributions from resonance-exchanges 
through the $\pi{\cal A} NN$ vertex 
using the estimates of Ref.\cite{meissner}, of which relevant diagrams are
illustrated in Fig.~2.
We have separated the zero-range part of the OPE contribution
from the finite-range part. 
The former is subject to the short-range correlation (SRC)
between nucleons which we will discuss below.
Note that the last column is obtained taking the values of $c_3$ and $c_4$ as
determined from the experiment in Ref.\cite{meissner}; 
it is not just the sum of the contributions from
the resonance-exchanges and the kinetic term.
The ${\it naive}$ quenching of the axial-vector coupling due to the 
OPE graph (Fig.~1$(a)$) is by the amount of $\delta g_A/g_A\simeq -0.52$.
In order to reproduce the experimental coupling
constant $g_A^*\simeq 1.0$, 
the four-fermion contact graph (Fig.~1$(b)$) needs to enhance the
coupling by about $0.32$.
Meanwhile, the four-nucleon contact interaction contains 
implicitly the contribution of Fig.~3, 
which is the main part of the conventional 
mechanism of the quenching.
This contradictory behavior, however, disappears when we incorporate the
SRC. 
\begin{figure}
\centerline{\epsfig{file=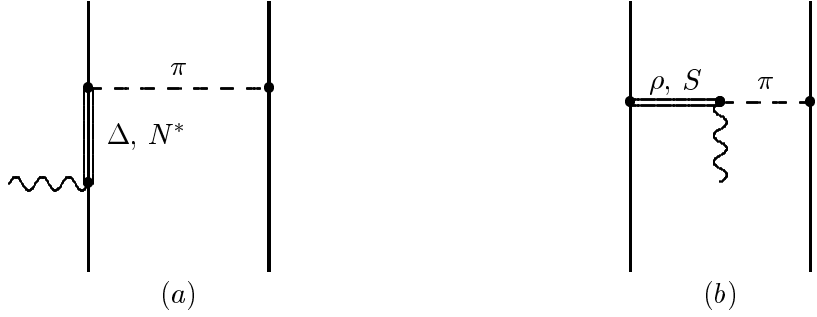}}
\caption{
The resonance-exchange graphs which contribute to Fig.~1$(a)$.
}
\label{GTO}
\end{figure}
\begin{figure}
\centerline{\epsfig{file=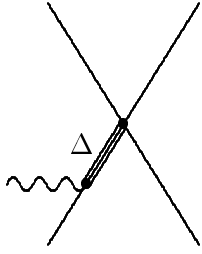}}
\caption{
A graph contributing to Fig.~1$(b)$.
}
\label{SRC}
\end{figure}

For the purpose of including the SRC,
it is much  more convenient to work in the position space.
In regard to  the SRC,  it is worth paying
attention to the conventional wisdom of applying the nuclear correlation 
only to the exchange (Fock) channel.
The direct (Hartree) contribution is effectively from the 
expectation value of one-body operators.
Therefore the Hartree contribution is the same as eq.(\ref{gAHartree}).
Incorporating the SRC for the Fock term by cutting-off the
contribution from $r < r_c$,
we have 
\be
\frac{\delta^F g_A}{g_A} =
        \frac{m_\pi^3 \rho}{m_N f_{\pi}^2} 
 \int_{r_c}^\infty \! dr\, 
 \frac{r\, j_1(k_F r)}{k_F} \left[
\frac{k_F}{2 m_\pi} j_1(k_F r) Y_1(m_\pi r) 
- \left(\hatc_3- 2\hatc_4 + \frac14\right) j_0(k_F r) Y_0(m_\pi r)\right]
\label{gAcoII}\ee
where 
$j_0(x)= \frac{\sin(x)}{x}$, 
$j_1(x)= \frac{\sin(x)}{x^2} - \frac{\cos(x)}{x}$,
$Y_0(x)= \frac{\e^{-x}}{x}$ and $Y_1(x) = (1 + \frac{1}{x}) Y_0(x)$.
In Table 2, the Fock contribution is calculated for 
several values of $r_c$ and the Hartree 
contribution is adjusted to give 
the empirical quenching of $\delta g_A/g_A\simeq -0.2$.
We find that the OPE contribution
is of sizable importance, {\it i.e.},
$\frac{\delta g_A}{g_A} \simeq (0.09 \sim 0.15)$.
Nevertheless, the quenching comes mainly from the four-nucleon
contact interaction.
The importance of the short-range interaction in the quenching of the
axial-vector coupling is a well-known observation in the conventional
approaches. In this work, such an observation is naturally understood 
based on  chiral perturbation theory. 

The density dependence of $\delta g_A/g_A$ with $r_c=0^+$ has been plotted
in Fig.~4.
The leading
non-trivial medium correction is roughly linear in the density and
this aspect is mainly due to the fact that the zero-range 
interaction accounts for the quenching for the most part.

For comparison with the conventional approach, 
we estimate the corresponding Landau-Migdal parameter $g_{N\Delta}'$ in
the effective $NN\leftrightarrow N\Delta$ potential by 
identifying the potential
from the four-nucleon contact  interaction with the phenomenological one as
in Ref.\cite{rho1}:
\be
\hat d_1 = \frac{4 m_N f_\pi^2 f_\Delta^2}{9 (m_\Delta-m_N) m_\pi^2} 
g_{N\Delta}'.
\ee
Our result shown in Table~2, $g_{N\Delta}'=(0.37 \sim 0.43)$, 
lies well within the range of the empirical value
with $\frac{f_\Delta^2}{4\pi} \simeq 0.32$\cite{theory}.

\begin{table}
\begin{center}
\begin{tabular}{|l|c|c||c|c|}\hline
   & Fock (OPE) & Hartree (Contact) & ${\hat d}_1$  & $g_{N\Delta}'$\\ \hline
Without OPE&                 $0$    & $-0.21$ & $0.64$ & $0.25$ \\ \hline
 $(r_c=0^+)$ &                  $0.15$ & $-0.36$ & $1.11$ & $0.43$ \\
 $(r_c=0.4\ \mbox{fm})$ &     $0.13$ & $-0.34$ & $1.04$ & $0.40$ \\
 $(r_c=0.7\ \mbox{fm})$ &     $0.09$ & $-0.30$ & $0.94$ & $0.37$ \\ \hline
\end{tabular}
\end{center}
\caption\protect{
Fock (2nd column) and Hartree (3rd column) contributions to $\delta g_A/g_A$
at normal nuclear density with various cut-offs. 
The ${\hat d}_1$ is fitted to give 
$\delta g_A \simeq -0.26$.
In the second row, we have also listed the result without OPE.
}
\end{table}

\begin{figure}
\centerline{\epsfig{file=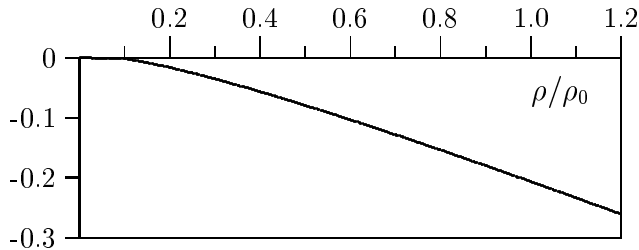}}
\caption{
$\delta g_A/g_A$ with respect to the density. 
The cut-off is taken as $r_c=0^+$.
}
\label{xA}
\end{figure}

In more realistic calculations, surface effects as well as 
short-range correlations between nucleons will be incorporated in the 
wavefunction.   At present, such a wavefunction
for nuclear matter or finite nuclei has not been obtained using
chiral perturbation theory. Thus, it would be interesting to study how
the above features in the independent particle approximation change
in more realistic approximations.
Furthermore we believe it worth investigating 
the nuclear medium modification of the physical quantities such as 
the nucleon mass and the pion decay constant in relation
with the ``Brown-Rho scaling''\cite{BR} in the light of present work.

\vspace{1 in}
\centerline{\bf ACKNOWLEDGMENTS} 
\vspace{0.3 in}
We are very grateful to Mannque Rho for valuable comments.
This work was supported in part by Korea Ministry of Education, 
Korea (BSRI-96-2418, BSRI-96-2441) and in part 
by Korea Science and Engineering Foundation through 
Center for Theoretical Physics, SNU.


\begin{thebibliography}{50}

\bibitem{exp}
B. H. Wildenthal, M. S. Curtin and B. A. Brown,
 Phys. Rev. {\bf C28} (1983) 1343;
D. H. Wilkinson in ``{\it Physics with Heavy Ions and Mesons}",
ed. R. Balian, M. Rho and G. Ripka (North Holland, Amsterdam, 1978);
B. Buck and S. M. Perez, \prl\ {\bf 50} (1983) 1975;
K. Langanke, D. J. Dean, P. B. Radha, Y. Alhassid and S. E. Koonin,
``{\it Shell-model Monte Carlo studies of fp-shell nuclei}",
nucl-th/9504019.

\bibitem{theory}
See {\it Pions and Nuclei} by T. Ericson and W. Weise 
(Clarendon, Oxford, 1988) and 
references therein. 

\bibitem{pioncond} {\it Mesons in Nuclei}, 
 ed. M. Rho and D. Wilkinson (North Holland,
 Amsterdam--New York--Oxford, 1979).

\bibitem{rho1}
K. Ohta and M. Wakamatsu, \np\ {\bf A234} (1974) 445;
M. Rho, Nucl. Phys. {\bf A231} (1974) 493.

\bibitem{BR} G.E. Brown and M. Rho, 
Phys. Rev. Lett. {\bf 66} (1991) 2720;
M. Rho, J. Korean Phys. Soc. {\bf 29} (1996) S337;
B. Friman and M. Rho, \np\ {\bf A606} (1996) 303.


\bibitem{Weinberg}
S. Weinberg, Phys. Lett. {\bf B251} (1990) 288; 
Nucl. Phys. {\bf B363} (1991) 3;
Phys. Lett. {\bf B295} (1992) 114.

\bibitem{pmr1}
T-S. Park,\ D.-P. Min and M. Rho,\ Phys. Rep. {\bf 233}
(1993) 341.

\bibitem{Kolck1}
C. Ordonez, L. Ray, and U. van Kolck, Phys. Rev. Lett. {\bf 72} (1994) 1982; 
Phys. Rev. C{\bf 53} (1996) 2086;
U. van Kolck, Phys. Rev. C{\bf 49} (1994) 2932.
\bibitem{KDR} K. Kubodera, J. Delorme and M. Rho,
Phys. Rev. Lett. {\bf 40} (1978) 755.
\bibitem{TSP} T.-S. Park, D.-P. Min and M. Rho,
\prl\ 74 (1995) 4153;
T.-S. Park, D.-P. Min and M. Rho, \np\ {\bf A 596} (1996) 515;
T.-S. Park, I. S. Towner and K. Kubodera, \np \ {\bf A579} (1994) 381.

\bibitem{rho2}
M. Rho, Phys. Rev. Lett. {\bf 66} (1991) 1275

\bibitem{meissner}
V. Bernard, N. Kaiser and Ulf-G. Meissner,
Nucl. Phys. {\bf A615} (1997) 483.

\bibitem{Kolck3} T. D. Cohen, J. L. Friar, G. A. Miller
and U. van Kolck, Phys. Rev. {\bf C53} (1996) 2661.

\end{thebibliography}
\end{document}